\documentclass[a4paper,11pt]{article}

\usepackage{jheppub}
\usepackage{graphicx}
\usepackage{amsmath}
\usepackage{placeins}

\definecolor{mycolor}{rgb}{0.984375, 0.398438, 0.}
\hypersetup{
    pdftitle={The Singular Structure of Rapidity Distributions for Color-Singlet Production at N3LO},
    colorlinks=true,
    linkcolor=mycolor,
    citecolor=mycolor,
    filecolor=mycolor,      
    urlcolor=mycolor
}

\newcommand{\refcite}[1]{ref.~\cite{#1}}
\newcommand{\refscite}[1]{refs.~\cite{#1}}
\newcommand{\eq}[1]{eq.~\eqref{eq:#1}}
\newcommand{\eqs}[2]{eqs.~\eqref{eq:#1} and \eqref{eq:#2}}
\renewcommand{\sec}[1]{section~\ref{sec:#1}}
\newcommand{\app}[1]{appendix~\ref{app:#1}}

\newcommand{\df}{\mathrm{d}}
\newcommand{\eps}{\epsilon}

\newcommand{\cI}{\mathcal{I}}
\newcommand{\cL}{\mathcal{L}}
\newcommand{\cO}{\mathcal{O}}
\newcommand{\nlim}{\lim\limits_{\mathrm{strict}~n-\mathrm{coll.}}}
\newcommand{\as}{\alpha_s}
\newcommand{\nn}{\nonumber}
\newcommand{\MSbar}{\overline{\mathrm{MS}}}

\preprint{CERN-TH-2026-242, SLAC-PUB-260930}

\title{The Singular Structure of Rapidity Distributions for Color-Singlet Production at N$^3$LO}

\author[a]{Federica Devoto,}
\emailAdd{federica@slac.stanford.edu}

\author[a]{Bernhard Mistlberger,}
\emailAdd{bernhard.mistlberger@gmail.com}

\author[b]{and Gherardo Vita}
\emailAdd{gherardo.vita@cern.ch}

\affiliation[a]{SLAC National Accelerator Laboratory, Stanford University, Stanford, CA 94039, USA}

\affiliation[b]{CERN, Theoretical Physics Department, CH-1211 Geneva 23, Switzerland}

\abstract{
We determine the universal singular structure of rapidity distributions for color-singlet production in hadron collisions through next-to-next-to-next-to-leading order (N$^3$LO) in QCD. 
Our result captures the full singular contributions as either partonic momentum fraction approaches one, including the delta-function terms, while retaining the exact dependence on the other.
The essential new ingredient is the complete calculation of the three-loop quark and gluon rapidity beam-function matching kernels. 
Our calculation is based on the strict collinear expansion of the partonic cross sections for Higgs boson and Drell--Yan production.
We represent our results in terms of generalized polylogarithms as well as numerical series expansions.
The resulting N$^3$LO partonic coefficient functions apply to arbitrary color-singlet production processes with the process dependence entering through the hard function, which encodes the virtual corrections to the Born process.
}

\begin{document}
\maketitle

\section{Introduction}
\label{sec:intro}

The increasing precision of measurements at the Large Hadron Collider (LHC),
together with the much larger data set anticipated at the High-Luminosity
LHC, makes percent-level perturbative control a central requirement for
collider phenomenology~\cite{ATLAS:2019mfr}.
This target is not generically guaranteed by a next-to-next-to-leading order
(NNLO) calculation: for benchmark Higgs and Drell--Yan observables, the
next-to-next-to-next-to-leading order (N$^3$LO) corrections themselves
can induce percent-level or larger shifts, and in some fiducial regions expose
poor perturbative convergence through NNLO~\cite{Mistlberger:2018etf,
Chen:2022cgv,Gehrmann-DeRidder:2026edl}.
Third-order information is therefore needed to assess the perturbative series
at the precision demanded by current and future measurements.
Inclusive cross sections at N$^3$LO in QCD are available for Higgs production in gluon fusion,
Drell--Yan production, vector-boson-fusion Higgs production in the
structure-function approximation, and associated production of a Higgs boson
with a vector boson ($VH$)~\cite{Anastasiou:2015vya,Mistlberger:2018etf,Duhr:2020sdp,Duhr:2020kzd,Duhr:2019kwi,Duhr:2020seh,Duhr:2021vwj}.
Rapidity distributions and fully differential predictions have followed for
Higgs boson and neutral- and charged-current Drell--Yan
production~\cite{Dulat:2018bfe,Chen:2021isd,Chen:2021vtu,Chen:2022cgv,
Chen:2022lwc}, and, most recently, for boosted $VH$ production~\cite{Gehrmann-DeRidder:2026edl}.

These advances also expose a practical limitation: exact fully differential
N$^3$LO calculations can require millions of core hours~\cite{Gehrmann-DeRidder:2026edl},
making exhaustive studies of scale and parton distribution function (PDF) dependence, fiducial selections, and collider energies prohibitively
expensive.
The need for fast reevaluation is particularly acute, for example, in global PDF
determinations, where Drell--Yan data provide essential flavor constraints
and the theory prediction must be recomputed many times during a fit.
Current approximate-N$^3$LO fits combine incomplete hard-scattering
information with approximations or $K$-factor
treatments~\cite{NNPDF:2024nan,McGowan:2022nag}, while full N$^3$LO
interpolation grids remain beyond reach because of their computational
cost~\cite{Cruz-Martinez:2025ffa}.
This motivates systematically improvable approximations that retain the
dominant higher-order corrections while remaining inexpensive enough for
large-scale phenomenological applications.

Threshold expansions provide a classical example of this strategy.
They exploit the universal structure of soft radiation near partonic
threshold~\cite{Sterman:1986aj,Catani:1989ne}, but by construction discard
radiation that is energetic and collinear to either incoming hadron.
At leading power, the threshold approximation retains only the partonic
channels connected to the Born process by soft radiation.
This limitation is especially acute for quark-initiated processes: channels
involving an initial-state gluon first enter at subleading power, but can be
phenomenologically enhanced by the large gluon luminosity at LHC
energies~\cite{Mistlberger:2025lee}.
For gluon-initiated production, by contrast, the approximation at least retains
the gluon-gluon channel, which is often dominant.
Leading-power threshold results alone are therefore not a sufficiently general
basis for precision phenomenology.

One way to extend their reach is to calculate successive power corrections
about threshold.
This strategy was developed to high orders in the N$^3$LO calculation of the
inclusive gluon-fusion Higgs cross section, for which dozens of subleading
terms were retained~\cite{Anastasiou:2013srw,Anastasiou:2014lda,
Anastasiou:2015yha,Anastasiou:2015vya}, and was subsequently applied to
differential Higgs production and to its rapidity distribution through six
terms in the threshold expansion~\cite{Dulat:2017prg,Dulat:2018bfe}.
A complementary development is to preserve the exact dependence on one
incoming momentum fraction while expanding in the other.
Expansions around this limit have been developed in~\refscite{Lustermans:2019cau,Ebert:2020lxs,Mistlberger:2025lee}.
This collinear or \emph{generalized threshold} limit naturally contains the
threshold limit. At leading power, it captures an infinite tower of terms in
the threshold expansion for the diagonal channels and also describes singly
off-diagonal channels such as $qg$ in color-singlet production.
The collinear approximation of \refcite{Mistlberger:2025lee} combines the two
single-collinear limits in a factorization theorem with a convolution structure that generates contributions that are
formally subleading in both limits, including doubly off-diagonal channels such as $gg$ in Drell--Yan or diphoton production.
At NNLO, the collinear approximation accurately reproduces the Higgs boson and
Drell--Yan rapidity distributions throughout the phenomenologically relevant
range.
Its improvement over the standard threshold approximation is particularly
striking for electroweak-scale Drell--Yan production, where it captures the
sizable and strongly anticorrelated $q\bar q$ and $qg$ contributions, including
the delicate cancellation that leaves a very small net NNLO
correction~\cite{Mistlberger:2025lee}.
Once its universal ingredients are known, evaluating the approximation is
inexpensive enough for laptop-level phenomenological studies.

This approximation relies on a factorization theorem which separates hard, soft, and collinear dynamics.
Its process dependence resides in the hard function, which encodes the virtual
corrections to the underlying Born process.
For Drell--Yan and gluon-fusion Higgs production in the heavy-top effective
theory, the relevant quark and gluon form factors are known through four
loops~\cite{Lee:2022nhh}.
For diphoton production, they are known through three loops~\cite{Caola:2020dfu,Bargiela:2021wuy}.
Soft radiation is encoded in the rapidity soft function~\cite{Mistlberger:2025lee}.
It is completely determined by the inclusive threshold soft function and is
therefore known through three loops~\cite{Anastasiou:2014vaa,Li:2014afw}.

The universal ingredients that encode collinear radiation in this
factorization theorem are the rapidity beam functions~\cite{Mistlberger:2025lee}, referred to as
generalized-threshold beam functions in \refcite{Lustermans:2019cau}.
For perturbative values of their measurement variable, they can be matched
onto standard light-cone PDFs.
The corresponding matching kernels were first obtained through NNLO in
\refcite{Lustermans:2019cau} by integrating over the fully differential quark
and gluon beam functions~\cite{Jain:2011iu,Gaunt:2014xxa,Gaunt:2020xlc} with the addition of a specific measurement constraint.
Although three-loop matching kernels are known for $N$-jettiness and
transverse-momentum-dependent beam functions~\cite{Ebert:2020unb,Baranowski:2022vcn,Ebert:2020yqt,Luo:2019szz,Luo:2020epw},
the rapidity beam-function kernels have remained unknown.
Moreover, the fully differential beam functions entering the original
representation are themselves known only through NNLO, and their direct extension
to three loops would require retaining additional measurement dependence that
is presently out of reach.
A crucial development in \refcite{Mistlberger:2025lee} was a direct definition
of the rapidity beam-function matching kernels as one-dimensional measurements
of strict collinear limits of physical color-singlet partonic cross sections.
This construction bypasses the fully differential beam functions and was used
in that work to rederive the complete NNLO kernels.
It is the essential starting point for the present calculation and makes the
determination at N$^3$LO tractable.

In this article, we calculate these matching kernels through N$^3$LO for both quark and gluon beam functions and for all partonic channels.
Our computation proceeds directly from strict collinear limits of the partonic Higgs boson and Drell--Yan production cross sections, using the expansion framework of \refcite{Ebert:2020lxs}.

The analytic structure of the rapidity beam-function matching kernels is unusually rich.
At intermediate stages, the integration alphabet contains an algebraic letter involving the square root of a quartic polynomial.
Ultimately, this complication drops out, and we express our final result in terms of generalized polylogarithms~\cite{Goncharov:2001iea}.
We provide the complete renormalized kernels, as well as the partonic coefficient functions required for the collinear approximation, in electronic form together with this article.
Combined with the hard and rapidity soft functions, they supply the final universal ingredient for constructing the collinear approximation at N$^3$LO.

Our result is also related to threshold resummation at fixed partonic
rapidity, recently formulated in \refcite{DeRos:2026bcv}.
In the notation of that framework, our kernels determine the missing
three-loop single-soft constant, i.e., the three-loop coefficient $g_{0,3}$ of
the function $g_0(\alpha_s)$.
The translation between the beam-function matching kernels and this
coefficient is in principle straightforward but algebraically cumbersome.
It depends on how lower-order matching coefficients, anomalous dimensions, and
the known three-loop soft-function constants are distributed between the
exponent and the prefactor of the resummation formula, and on the space in
which it is formulated.
It is also affected by the process dependence of the hard function.
This exercise was carried out at one loop for Drell--Yan production in
\refcite{DeRos:2026bcv}, translating the one-loop rapidity beam function of
\refcite{Lustermans:2019cau} into $g_{0,1}$.

This paper is organized as follows.
In \sec{coll_exp}, we define the rapidity beam function and explain its
extraction from the strict collinear limit of color-singlet production cross
sections.
In \sec{results}, we describe the calculation and analytic structure of the matching kernels, their numerical representations, the construction of the universal partonic coefficient functions, and our validation checks.
We conclude in \sec{conclusion}.
Our renormalization conventions are collected in \app{details}.

\section{Rapidity beam functions from the collinear limit of cross sections}
\label{sec:coll_exp}

We consider the inclusive production of a color-singlet final state $h$ in the
collision of two hadrons,
\begin{align}
 P(P_1)+P(P_2)\to h(p_h)+X(k)\,.
\label{eq:hadronic_process}
\end{align}
The total momentum of the hadronic radiation is denoted by $k$, and we define
\begin{align}
 S=(P_1+P_2)^2\,,\qquad
 Q^2=p_h^2\,,\qquad
 \tau=\frac{Q^2}{S}\,,\qquad
 Y=\frac12\ln\frac{P_2\cdot p_h}{P_1\cdot p_h}\,.
\label{eq:hadronic_variables}
\end{align}
It is convenient to introduce the Born momentum fractions
\begin{align}
 \xi_1=\sqrt{\tau}\,e^{Y}\,,\qquad
 \xi_2=\sqrt{\tau}\,e^{-Y}\,,\qquad
 \bar\xi_i=1-\xi_i\,.
\label{eq:xi_variables}
\end{align}
The hadronic rapidity distribution can then be written as
\begin{align}
 Q^2\frac{\df\sigma}{\df Q^2\df Y}
 =\tau\sum_{i,j}
 f_i(\xi_1,\mu^2)\otimes_{\xi_1}
 \eta_{ij}(\xi_1,\xi_2,\mu^2)
 \otimes_{\xi_2}f_j(\xi_2,\mu^2)\,,
\label{eq:hadronic_rapidity}
\end{align}
where $\eta_{ij}$ is the partonic coefficient function and
\begin{align}
 f(x)\otimes_x g(x)
 =\int_x^1\frac{\df z}{z}\,f(z)g\left(\frac{x}{z}\right)
\label{eq:mellin_convolution}
\end{align}
denotes a Mellin convolution.
We work in dimensional regularization with $d=4-2\eps$ and renormalize in the
$\MSbar$ scheme.

The limit $\xi_2\to1$ forces all additional radiation to become soft or
collinear along the beam direction selected by the limit.
At leading power in $\bar\xi_2$, the rapidity distribution factorizes as~\cite{Lustermans:2019cau,Ebert:2020lxs,Mistlberger:2025lee}
\begin{align}
 Q^2\frac{\df\sigma}{\df Q^2\df Y}
 =\tau\sum_{i,j}H_{ij}(Q^2)\,
 B_i^Y(\bar \xi_1;\bar\xi_2)\otimes_{\xi_2}f_j(\xi_2)
 +\cO(\bar\xi_2^0)\,.
\label{eq:single_collinear_factorization}
\end{align}
For the moment all quantities in \eq{single_collinear_factorization} are bare.
The hard function $H_{ij}$ contains virtual corrections to the underlying
Born process, whereas the rapidity beam function $B_i^Y$ describes the soft-collinear and energetic collinear radiation in the unresolved final state.
The argument to the right of the semicolon is its measurement variable (the small quantity), while the
argument to the left is the complement of the usual longitudinal momentum fraction, an $\cO(1)$ quantity that is not expanded.

When the characteristic virtuality $Q^2\bar\xi_2 \sim \mu_B^2$ is perturbative, the rapidity beam function admits an operator product expansion onto standard PDFs,
\begin{align}
 B_i^Y(\bar \xi_1,\mu^2;\bar\xi_2)
 =\sum_j \int_{\xi_1}^1\frac{\df z}{z}\,
 I_{ij}^Y(\bar z,\mu^2;\bar\xi_2)\,
 f_j\!\left(\frac{\xi_1}{z},\mu^2\right)\,.
\label{eq:beam_matching}
\end{align}
Here $\xi_1$ is the hadronic Born momentum fraction fixed by $Q^2$, $S$, and $Y$ through \eq{xi_variables}.
The perturbative matching kernel $I_{ij}^Y$ is the object calculated in this
article.
To relate it to a partonic cross section, we consider
\begin{align}
 i(p_1)+j(p_2)\to h(-p_h)+X_n(-p_3,\ldots,-p_{n+2})
\label{eq:partonic_process}
\end{align}
and denote the total momentum of the radiation, in the convention where all momenta are incoming, by
$k=-p_1-p_2-p_h$.
Following \refcite{Ebert:2020lxs}, we parameterize the radiative phase space by
\begin{align}
 w_1=-\frac{2p_1\cdot k}{2p_1\cdot p_2}\,,\qquad
 w_2=-\frac{2p_2\cdot k}{2p_1\cdot p_2}\,,\qquad
 x=\frac{k^2}{s\,w_1w_2}\,,
\qquad s=(p_1+p_2)^2\,.
\label{eq:partonic_variables}
\end{align}
In terms of the partonic coefficient function differential in these variables, the bare matching kernel is given by the one-dimensional measurement of the strict collinear limit
\begin{align}
 I_{ij}^{Y,\mathrm{bare}}(\bar z;\bar \xi_2)
 &=
 \int_0^1\df x\int_0^\infty\df w_1\,\df w_2\,
 \delta\bigl(w_2-\bar z\bigr)
 \nn\\
 &\quad\times
 \nlim\left\{
 \delta\left(
 \bar \xi_2-
 \frac{2-w_2-w_2x}{2(1-w_2)}w_1
 \right)
Q^2  \frac{\df\eta_{j\bar i}}
 {\df Q^2\,\df w_1\,\df w_2\,\df x}
 \right\}.
\label{eq:kernel_from_collinear_limit}
\end{align}
Here $\bar i$ denotes the flavor conjugate to $i$ in the crossed partonic
channel, and $\bar z=1-z$ denotes the complementary partonic momentum fraction. When the kernel is inserted into \eq{beam_matching}, its
longitudinal variable is identified as $z\to\xi_1$.
Equation~\eqref{eq:kernel_from_collinear_limit} is the crucial link between the rapidity beam function and the strict collinear expansion of a physical cross section.
Unlike the definition of the kernel as a projection of a fully differential beam
function \cite{Lustermans:2019cau}, \eq{kernel_from_collinear_limit} requires only a single nontrivial measurement constraint after
the collinear limit has been taken.

The two single-collinear limits can be combined without double counting their soft overlap.
Introducing the inverse of the rapidity soft function, $S_{r_i}^{-1}$, the resulting
factorization theorem can be expressed as \cite{Mistlberger:2025lee}
\begin{align}
 Q^2\frac{\df\sigma}{\df Q^2\df Y}
 &=\tau\sum_{i,j}H_{ij}(Q^2,\mu^2)
 \left[
 B_i^Y(\bar \xi_1,\mu^2;\bar\xi_2)
 \otimes_{\xi_1,\xi_2}
 S_{r_{i}}^{-1}(\mu^2;\bar\xi_1,\bar\xi_2)
 \right.
 \nn\\[-1mm]
 &\hspace{5.0cm}\left.
 {}\otimes_{\xi_1,\xi_2}
 B_j^Y(\bar \xi_2,\mu^2;\bar\xi_1)
 \right]
 +\cO(\bar\xi_1^0,\bar\xi_2^0)\,.
\label{eq:double_collinear_factorization}
\end{align}
We refer to the above formula as the collinear approximation to the hadronic cross section.
The representation $r_{i}$ is fundamental for a quark-antiquark Born
channel and adjoint for a gluon-gluon Born channel.
The double Mellin convolution appearing here is
\begin{align}
 f(x_1,x_2)\otimes_{x_1,x_2}g(x_1,x_2)
 =\int_{x_1}^1\frac{\df z_1}{z_1}
\int_{x_2}^1\frac{\df z_2}{z_2}
 f(z_1,z_2)
 g\left(\frac{x_1}{z_1},\frac{x_2}{z_2} \right).
\label{eq:double_mellin_convolution}
\end{align}
Note that in ref.~\cite{Mistlberger:2025lee} the factorization theorem was defined using a mixed Mellin--Laplace convolution, which gives a formally equivalent result up to terms beyond the validity of the approximation, i.e.\ terms that are regular in both $\bar\xi_1$ and $\bar\xi_2$. 
We change the definition here to homogenize the convolution structure, at the expense of a slightly more involved implementation of its resummation, which we do not pursue in this article. See ref.~\cite{Mistlberger:2025lee} for details.

The inverse rapidity soft function removes the overlap of the two collinear sectors.
The soft function itself is obtained from the eikonal limit of
the diagonal matching kernel,
\begin{align}
 S_{r_i}(\mu^2;\bar\xi_1,\bar\xi_2)
 =\lim_{\bar z\to0}I_{ii}^Y(\bar z,\mu^2 ;\bar\xi_2)\Big|_{\bar z=\bar\xi_1}\,,
\label{eq:soft_from_beam}
\end{align}
where the limit denotes the leading-power term as $\bar z\to0$, i.e.\ the terms proportional to $\delta(\bar z)$ and to plus distributions in $\bar z$.
The soft function is symmetric in its two arguments, $S^{r}(\mu^2;\bar\xi_1,\bar\xi_2)=S^{r}(\mu^2;\bar\xi_2,\bar\xi_1)$.
Its convolution inverse is defined by
\begin{align}
 S_r^{-1}  \otimes_{\xi_1,\xi_2} S_r
 =\delta(\bar\xi_1)\delta(\bar\xi_2)\,.
\label{eq:soft_convolution_inverse}
\end{align}

The bare kernel contains ultraviolet and infrared poles.
Coupling renormalization, the ultraviolet renormalization of the
soft-collinear operator, and PDF mass factorization give the finite kernel
\begin{align}
 I_{ij}^Y(\bar z,\mu^2;\bar{\xi_2})
 &=
 Z_H^{r_i,-1}(Q^2,\mu^2)
 \sum_k
 \Gamma_s^{r_i}(1-\bar{\xi_2},\mu^2)
 \circ_{\bar{\xi_2}}
 \left[
 {\bf Z}_{\as}\,
 I_{ik}^{Y,\mathrm{bare}}(\bar z;\bar{\xi_2})
 \right]
 \otimes_z\Gamma_{kj}(z,\mu^2)\,.
\label{eq:kernel_renormalization}
\end{align}
In the above equation, ${\bf Z}_{\as}$ denotes $\MSbar$ coupling renormalization,
\begin{equation}
\label{eq:coupling_renormalization_FD}
\as^{\mathrm{bare}}
 = \as(\mu^2)\left(\frac{\mu^2e^{\gamma_E}}{4\pi}\right)^{\eps} Z_{\as}(\mu^2)\,,\qquad
Z_{\as} = 1-\frac{\beta_0}{\eps}\frac{\as}{4\pi}
 +\Big(\frac{\beta_0^2}{\eps^2}-\frac{\beta_1}{2\eps}\Big)\Big(\frac{\as}{4\pi}\Big)^2
 +\cO(\as^3)\,,
\end{equation}
with $\beta_0$ and $\beta_1$ given in \eq{beta_coefficients},
and ${\bf Z}_{\as} I^{Y,\mathrm{bare}}$ is shorthand for expressing the bare kernel in terms of $\as(\mu^2)$ via this relation and re-expanding.
Moreover, $Z_H^{r_i}$ is the $\MSbar$ counterterm of the hard function, $H_{ij}(Q^2,\mu^2) = Z_H^{r_i}(Q^2,\mu^2,\eps)\,H_{ij}^{\mathrm{bare}}(Q^2,\eps)$. 
This counterterm is tied, via \eq{ZH_all_order}, to the non-cusp anomalous dimension of the quark or gluon form factor (sometimes referred as the \emph{collinear anomalous dimension}) which has been calculated to 4 loops in \refcite{Agarwal:2021zft}.
For the gluon channels, which we compute in the heavy-top effective theory, the hard function includes the Born factor $\as^2(\mu^2)$ and the Wilson coefficient of the effective operator, such that its anomalous dimension is the one given in \app{counter_term}.
Equivalently, in Laplace space the renormalization takes the form \eqref{eq:kernel_renormalization_laplace}, in which $Z_H^{r_i}$ enters only through the combination $\widetilde Z_B^{r_i}$ of \eq{ZB_definition}.

Our convention for the full PDF counterterm $\Gamma_{ij}$ is fixed by
\begin{align}
 f_i^{\mathrm{bare}}(x)
 &=
 \sum_j\Gamma_{ij}(x,\mu^2,\epsilon)
 \otimes_x f_j(x,\mu^2)\,,
\qquad
 f_i(x,\mu^2)
 =
 \sum_j\Gamma_{ij}^{-1}(x,\mu^2,\epsilon)
 \otimes_x f_j^{\mathrm{bare}}(x)\,.
\label{eq:pdf_counterterm_definition}
\end{align}
We suppress the dependence of the counterterms on the dimensional regulator
elsewhere.
The kernel $\Gamma_s^{r_i}$ is the leading endpoint contribution to the
diagonal component,
\begin{align}
 \Gamma_{i i}(z,\mu^2)
 =\Gamma_s^{r_i}(z,\mu^2)+\cO\bigl((1-z)^0\bigr)\,.
\label{eq:threshold_counterterm}
\end{align}
Here $\circ_{t}$ denotes the Laplace convolution in the measurement variable,
\begin{align}
 (f\circ_{t} g)(t)\equiv\int_0^t\df t'\, f(t')\,g(t-t')\,,
\label{eq:laplace_convolution}
\end{align}
which, in terms of $x=1-t$, is the leading term of the Mellin convolution in the expansion around $x\to 1$.
Explicit expressions and our perturbative conventions are collected in
\app{details}.

We expand the renormalized matching kernels as
\begin{align}
 I_{ij}^Y(\bar z,\mu^2;\bar\xi_2)
 =\sum_{n=0}^{\infty}
 \left(\frac{\as(\mu^2)}{\pi}\right)^n
 I_{ij}^{Y,(n)}(\bar z,\mu^2;\bar\xi_2)\,.
\label{eq:kernel_expansion}
\end{align}
In the ancillary files, $\texttt{x1b}=\bar z$.
Thus distributions in \texttt{x2b} act on the beam-function measurement, while the dependence on \texttt{x1b}, including its endpoint distributions, encodes the longitudinal momentum fraction.

\section{Results}
\label{sec:results}

\subsection{Beam function matching kernels}
Our calculation starts from the partonic cross sections for Higgs boson
production in gluon fusion and for the production of an off-shell photon in
the Drell--Yan process.
The Higgs boson cross section is evaluated in the heavy-top effective theory.
Together, the two processes give access to gluon- and quark-initiated beam
functions and to every independent flavor channel.
We expand the squared matrix elements and phase space in the strict collinear
limit before integration, following the method of
\refcite{Ebert:2020lxs}.
At N$^3$LO, the required contributions contain one, two, or three additional
partons in the final state and are conventionally denoted by RVV, RRV, and
RRR, respectively.
For the RVV contribution, we expand the available full-kinematics partonic
cross sections in the strict collinear
limit~\cite{Duhr:2020seh,Dulat:2014mda,Dulat:2017brz,Dulat:2017prg}.
These inputs build in part on the real--virtual Higgs contributions, the
two-loop splitting amplitudes, and the two-loop soft
current~\cite{Anastasiou:2013mca,Duhr:2014nda,Duhr:2013msa}.
For RRV and RRR, we expand the matrix elements before carrying out the loop
and phase-space integrations.

Loop and phase-space integrals are handled uniformly using reverse unitarity~\cite{Anastasiou2002,Anastasiou2003,Anastasiou:2002qz,Anastasiou:2003yy,Anastasiou2004a}.
We employ integration-by-parts identities~\cite{Tkachov1981,Chetyrkin1981,Laporta:2001dd} to reduce the expanded cross sections to
collinear master integrals.
We determine these integrals from differential equations~\cite{Kotikov:1990kg,Kotikov:1991hm,Kotikov:1991pm,Gehrmann:1999as,Henn:2013pwa}.
Boundary conditions are fixed in limits in which the collinear integrals
simplify, in particular by matching their further soft expansion to the
corresponding soft master
integrals~\cite{Duhr:2022cob,Anastasiou:2014lda,
Anastasiou:2014vaa,Anastasiou:2015yha,Duhr:2019kwi}.
After applying the measurement in
\eq{kernel_from_collinear_limit}, we renormalize the result according to
\eq{kernel_renormalization}.

At order $n$, the dependence on the measurement variable takes the form 
\begin{align}
 I_{ij}^{Y,(n)}(\bar z,\mu^2;\bar{\xi_2})
 &=
 \delta(\bar \xi_2)\,I_{ij}^{Y,(n,\delta)}(\bar z,\mu^2)
 +\sum_{m=0}^{2n-1}
 \cI_{ij}^{Y,(n,m)}(\bar z,\mu^2)\,
 \cL_m(\bar{\xi_2})\,,
\nn\\
 \cL_m(\bar{\xi_2})&=\left[\frac{\ln^m \bar{\xi_2}}{\bar{\xi_2}}\right]_+\,,
\label{eq:measurement_distribution}
\end{align}
where the coefficients are polynomials in
\begin{align}
 L_Q=\ln\frac{Q^2}{\mu^2}\,.
\label{eq:LQ}
\end{align}
They are distributions in the longitudinal momentum fraction and contain both ordinary functions and endpoint terms (delta functions and plus distributions) at $\bar z=0$.
All logarithmic scale dependence at three loops is constrained by the renormalization group equation in \app{RGE}.
The genuinely new information is the three-loop boundary condition $I_{ij}^{Y,(3,\delta)}(\bar z,\mu^2=Q^2)$. 
We nevertheless provide the complete kernels, including their full scale dependence, because this form is most convenient for direct applications.

There are seven independent matching kernels,
\begin{equation}
ij\in\{ gg,\,gq,\,qg,\,qq,\,q\bar q,\,q q^\prime,\, q \bar q^\prime\}.
\end{equation}
Here, $g$, $q$, $\bar q$, $q^\prime$, and $\bar q^\prime$ denote a gluon, a quark, an antiquark, a quark of a different flavor, and an antiquark of a different flavor, respectively.
The first two channels $\{gg,\,gq\}$ are derived from the Higgs boson cross section and the remaining ones from the Drell--Yan cross section.

The diagonal kernels reduce at tree level to
\begin{align}
 I_{ii}^{Y,(0)}(\bar z;\bar{\xi_2})
 =\delta(\bar z)\delta(\bar{\xi_2})\,,
\qquad i\in\{q,g\}\,.
\label{eq:tree_kernel}
\end{align}

The three-loop kernels contain generalized polylogarithms
(GPLs)~\cite{Goncharov:2001iea}.
They are defined by
\begin{align}
 G(a_1,\ldots,a_n;\bar z)
 =\int_0^{\bar z}\frac{\df y}{y-a_1}G(a_2,\ldots,a_n;y)\,,
 \qquad G(;\bar z)=1\,,
\label{eq:GPL_definition}
\end{align}
with the usual regularization when all trailing letters vanish.

When deriving canonical differential equations in the variable $\bar z$, we introduce three square roots:
\begin{align}
 r_1(\bar z)&=\sqrt{4-4\bar z-\bar z^2}\,,
\nn\\
 r_2(\bar z)&=\sqrt{7-2\bar z-\bar z^2}\,,
\nn\\
 r_3(\bar z)&=\sqrt{17-30\bar z+19\bar z^2-6\bar z^3+\bar z^4}\,.
\label{eq:algebraic_roots}
\end{align}
However, we find that after inserting our master integrals into the final matching kernels only the root $r_1$ survives.
To express it in terms of GPLs, we introduce the variable
\begin{equation}
s_4 =\frac{2 \bar z}{2-\bar z +r_1(\bar z)}\,,
\qquad
\bar z=\frac{4 s_4}{2+2 s_4+s_4^2}\,.
\end{equation}
This transformation rationalizes the square root,
\begin{equation}
r_1(\bar z)=\frac{2 \left(2-s_4^2\right)}{2+2 s_4+s_4^2}\,,
\end{equation}
and allows us to write all matching kernels in terms of GPLs.

The GPL alphabet required to express the functions without $r_1$ is given by
\begin{align}
 \mathcal A_R= \{\bar z-x_i\,|\,x_i\in S_R\},\hspace{1cm} S_R= \left\{-2,-1,0,\frac{1}{2},\frac{2}{3},1,\frac{4}{3},\frac{3}{2},2,3\right\},
\label{eq:GPL_alphabet}
\end{align}
and the one to express the functions with $r_1$ is given by
\begin{align}
 \mathcal A_{s_4}= \{s_4-x_i\,|\,x_i\in S_{s_4}\},\hspace{0.5cm} S_{s_4}= \left\{0,\,\pm1\pm i,\,\pm i\sqrt{2},\,-3\pm\sqrt{7}\right\}.
\label{eq:GPL_alphabets4}
\end{align}
The full alphabet is given by the union of $ \mathcal A_{s_4}$ and $ \mathcal A_{R}$.
Note that one of the zeros of $r_1^2$ lies inside the physical interval,
\begin{equation}
r_1^2=-(\bar z-q_1)(\bar z-q_2),\hspace{1cm} q_1=2\sqrt{2}-2\approx 0.828,\hspace{1cm} q_2=-2\sqrt{2}-2\approx -4.83.
\end{equation}
Consequently, the square root becomes imaginary for $\bar z>q_1$. Nevertheless, our matching kernels are physical quantities and therefore remain real. 
Moreover, they are smooth when crossing this point.
However, individual GPLs used to express our matching kernels carry an explicit imaginary part, which cancels in the full expression. To evaluate the GPLs consistently, an explicit continuation prescription is required, which in practice means equipping the variable $\bar z$ with an infinitesimally small positive imaginary part.

For the convenience of the user, we provide numerical series representations of the regular part (i.e.\ without distributions in $\bar z$) of our matching kernels. 
Specifically, we expand them to 180 terms around the limit $\bar z\to 0$ and to 120 terms around the limit $\bar z \to 1$. 
Matching the series expansions at $\bar z=\frac{3}{4}$ ensures a relative numerical accuracy better than $10^{-15}$ throughout the entire unit interval $\bar z\in[0,1)$. 

\begin{figure*}[h!]
 \centering
 \includegraphics[width=0.49\textwidth]{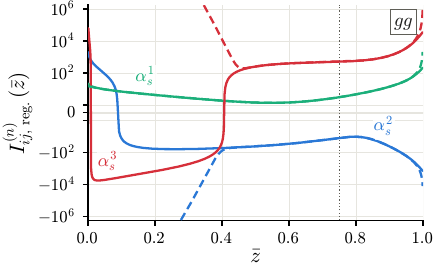}
 \hfill
 \includegraphics[width=0.49\textwidth]{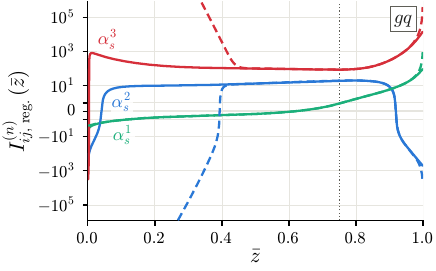}\\
 \includegraphics[width=0.49\textwidth]{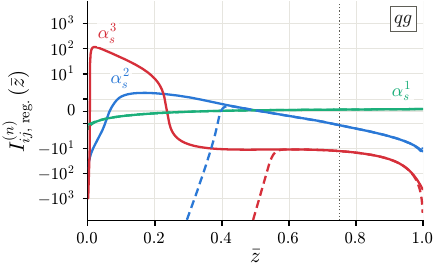}
 \hfill
 \includegraphics[width=0.49\textwidth]{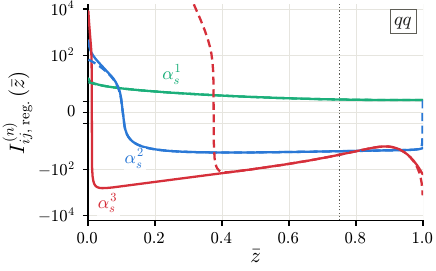}\\
 \includegraphics[width=0.49\textwidth]{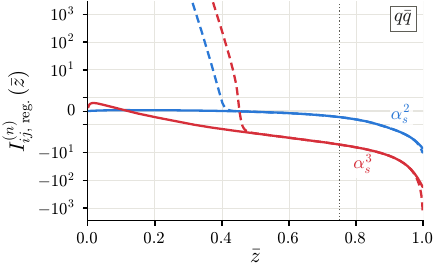}
 \hfill
 \includegraphics[width=0.49\textwidth]{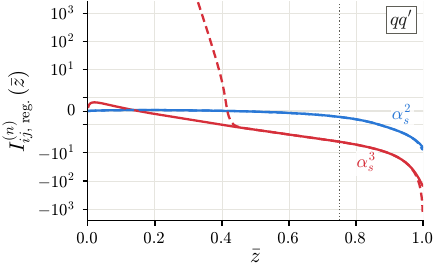}\\
 \includegraphics[width=0.49\textwidth]{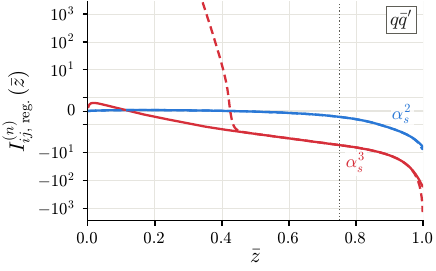}
 \hfill\mbox{}
 \caption{Regular part $I_{ij,\,\text{reg.}}^{(n)}(\bar z)$ of the $\delta(t)$ coefficient of the matching kernels at order $\alpha_s^n$ ($n=1,2,3$) for all channels $ij$, with $n_c=3$, $n_f=5$ and $\mu^2=Q^2$.
 Green, blue and red lines show the NLO, NNLO and N$^3$LO coefficients, respectively.
 Solid lines show the series expansion around $\bar z=0$, dashed lines the expansion around $\bar z=1$; the vertical dotted line marks the matching point $\bar z=3/4$.}
 \label{fig:Ireg}
\end{figure*}
Figure~\ref{fig:Ireg} shows the regular part of the $\delta(t)$ coefficient $I_{ij}^{Y,(n,\delta)}$ of all channels through N$^3$LO, evaluated with these series representations. Green, blue and red lines represent the results at NLO, NNLO and N$^3$LO respectively. 
The dashed lines represent the expansion around $\bar z\to1$ and the solid lines the expansion around $\bar z\to 0$. 
We confirmed the numerical results of our expansions by evaluating the GPLs explicitly with GiNaC~\cite{Bauer2000,Vollinga:2004sn} at several points.

\subsection{Rapidity approximation of the coefficient function} 
We analytically construct the rapidity approximation to the partonic coefficient function for color-singlet production at the LHC following \eq{double_collinear_factorization}.
Specifically, we define this quantity by
\begin{equation}
\label{eq:etadef}
\eta_{ij,\,kl}(\bar \xi_1,\bar \xi_2,\mu^2)= 
I_{ij}^{Y}(\bar \xi_1,\mu^2;\bar \xi_2)
 \otimes_{\xi_1,\xi_2} S^{r_{ik},-1}(\mu^2;\bar \xi_1,\bar \xi_2)
 \otimes_{\xi_1,\xi_2} I_{kl}^{Y}(\bar \xi_2,\mu^2;\bar \xi_1)\,.
\end{equation}
Combining the above formula with the hard function $H_{ik}$ describing a particular process, we compute the partonic cross section as follows
\begin{equation}
\eta_{jl}(\xi_1,\xi_2,\mu^2)=\sum_{i,k} H_{ik}(Q^2,\mu^2) \eta_{ij,\,kl}(\bar \xi_1,\bar \xi_2,\mu^2).
\end{equation}
Note that the above partonic cross section is differential in the virtuality and the rapidity of the colorless final state.
We relate it to the inclusive partonic cross section, expressed in terms of the variable $\hat\tau=Q^2/s$, via
\begin{equation}
\label{eq:increl}
\eta_{jl}(\hat\tau,\mu^2)= \int_0^1 \df \xi_1\df \xi_2\, \delta(\hat\tau-\xi_1 \xi_2)\, \eta_{jl}(\xi_1,\xi_2,\mu^2).
\end{equation}

Concrete results of this article are the convolutions of beam and soft functions that give the partonic coefficient functions $\eta_{ij,\,kl}(\bar \xi_1,\bar \xi_2,\mu^2)$ of \eq{etadef}. 
The collinear approximation is exact at leading power if either of the two variables approaches one, i.e.\ for $\xi_1\to1$ at arbitrary $\xi_2$ and for $\xi_2\to1$ at arbitrary $\xi_1$; the neglected terms are power suppressed in both limits simultaneously.
Using these functions naively to compute a hadronic cross section would yield predictions that are invariant under scale variations only up to such terms, which are beyond the formal accuracy of our approximation. 

To obtain a result that is exactly scale invariant, we derive the
dependence on $\mu^2$ from the evolution equations.
The partonic cross section is the product of the hard function and the
coefficient function, $H_{ik}\,\eta_{ij,\,kl}$.
Its scale dependence is fixed by the independence of the hadronic cross
section of $\mu^2$ and the evolution of the PDFs,
\eq{DGLAP}, while the scale dependence of the hard function is governed by the
cusp and collinear anomalous dimensions.
Together, they determine the evolution of the coefficient function,
\begin{align}
 \frac{\df}{\df\ln\mu^2}\,\eta_{ij,\,kl}
 =-\frac{\df\ln H_{ik}}{\df\ln\mu^2}\,\eta_{ij,\,kl}
 -\sum_m\Big[
 P_{mj}\otimes_{\xi_1}\eta_{im,\,kl}
 +\eta_{ij,\,km}\otimes_{\xi_2}P_{ml}
 \Big]\,,
\label{eq:eta_evolution}
\end{align}
where the derivatives also act on the argument of the strong coupling.
Expanding in $\as(\mu^2)$ and $L_Q$, this equation fixes all logarithmic terms
of $\eta_{ij,\,kl}$ at a given order in terms of the non-logarithmic terms of
the lower orders, the QCD $\beta$ function, the anomalous dimensions of the hard
function, and the splitting functions through three
loops~\cite{Moch:2004pa,Vogt:2004mw}.
We take the non-logarithmic terms from \eq{etadef} at $\mu^2=Q^2$ and solve
\eq{eta_evolution} through N$^3$LO.
The result differs from the direct evaluation of \eq{etadef} only by terms that
are regular in both $\bar\xi_1$ and $\bar\xi_2$, i.e.\ beyond the formal
accuracy of our approximation.

\subsection{Validation of results}
We perform several checks in order to validate the correctness of our results.
\begin{enumerate}
\item Infrared and ultraviolet singularities of our beam functions are removed by universal renormalization and mass factorization counterterms, see \eqs{kernel_renormalization}{pdf_counterterm_definition}. Indeed, all poles in the dimensional regulator up to $1/\eps^6$ in our bare matching kernels cancel, which provides a very stringent check.
\item We validate explicitly that the beam function matching kernels satisfy their renormalization group equation, \eq{kernel_RGE}.
\item The soft limit of the matching kernels corresponds to the rapidity soft function $S^{r_{i}}$, see \eq{soft_from_beam}, which can easily be derived from the inclusive soft function computed through N$^3$LO in refs.~\cite{Anastasiou:2014vaa,Li:2014afw}. We validate this correspondence. 
\item We verify that the first non-vanishing term in the limit $\bar z\to 0$ of the matching kernels corresponds to the matching kernel of the $N$-jettiness beam functions~\cite{Stewart:2009yx,Stewart:2010qs,Gaunt:2014xga,Gaunt:2014cfa,Ebert:2020unb,Baranowski:2022vcn}. This relation exists due to a correspondence of the partonic variables up to power corrections.
\item The collinear approximation to the cross section, \eq{double_collinear_factorization}, computes the soft and next-to-soft terms of the inclusive cross section~\cite{Lustermans:2019cau}. 
We use \eq{increl} to fully validate this against the analytic computation of the inclusive Higgs boson~\cite{Anastasiou:2014lda,Anastasiou:2015vya,Mistlberger:2018etf} and Drell--Yan~\cite{Duhr:2020seh,Duhr:2021vwj} cross sections at N$^3$LO.
\item In ref.~\cite{Dulat:2018bfe} one of the authors performed an analytic threshold expansion of the rapidity distribution of the gluon-fusion Higgs boson production cross section including the first six terms of this expansion. We validate that the first six terms of the expansion of our $I^{Y,(3)}_{gg}$ and $I^{Y,(3)}_{gq}$ matching kernels reproduce the collinear limit of this result.
\end{enumerate}

\section{Conclusions}
\label{sec:conclusion}

We have determined the universal singular structure of rapidity distributions
for color-singlet production in hadron collisions through N$^3$LO in QCD.
To achieve this result we applied the method of collinear expansion for LHC cross sections~\cite{Ebert:2020lxs} to the case of the rapidity measurement in Higgs boson
and Drell–Yan production at N$^3$LO and calculated the quark and gluon rapidity beam-function matching kernels in all partonic channels.
Combining the beam functions for each hadronic leg with the inverse rapidity soft function to remove the soft overlap of the two collinear expansions, we determine the universal coefficient functions, which we supply in analytic form in the ancillary files.
While these quantities have been extracted from the expansion of the Drell-Yan and Higgs production processes, they are universal quantities and can be used for any other color-singlet process by combining them with the appropriate hard functions.

The three-loop functions determined in this work involve a substantially richer function space than their lower-order counterparts.
In intermediate stages of the computation, we encounter iterated integrals over algebraic letters involving three square roots, including the square root of a quartic polynomial.
In our final result only functions involving a single square root of a quadratic polynomial are left, which we rationalize and express in terms of generalized polylogarithms.
We provide numerical series representations of the regular part of the kernels, accurate to better than $10^{-15}$ over the full range of the longitudinal momentum fraction, and show the coefficients of the delta function in the measurement variable in all seven independent channels.

We performed an extensive set of independent cross-checks. 
We verified the cancellation of poles in dimensional regularization, which at this order go up to $1/\eps^6$, as well as verifying that the logarithmically enhanced terms in the kernels correspond to those predicted by the renormalization group evolution. 
Moreover, we confirmed that the eikonal limit of our matching kernels matches the corresponding limit of the $N$-jettiness beam functions channel by channel, and that the one of the diagonal channels reproduces the rapidity soft function. 
Finally, at the cross-section level, we checked consistency with inclusive coefficient functions and the known threshold expansion of the Higgs rapidity distribution.
This work opens the way to fast N$^3$LO differential cross sections for color singlet processes.

\acknowledgments
We would like to thank Claude Duhr for fruitful discussions.
BM and FD are supported by the United States Department of Energy, Contract DE-AC02-76SF00515.
We thank the Erwin Schrödinger International Institute for Mathematics and Physics at the University of Vienna for partial support during the Programme ``New Paradigms for Harnessing Quantum Field Theory at Colliders'', July 27 -- August 28, 2026.

\appendix

\section{Renormalization of the rapidity beam-function matching kernels}
\label{app:details}

In this appendix we collect the renormalization group equations and
counterterms required to renormalize the bare rapidity beam-function matching
kernels through three loops.

\subsection{Renormalization group equations}
\label{app:RGE}

We define the Laplace transform with respect to the measurement variable by
\begin{align}
 \widetilde I_{ij}^Y(\bar z,\mu^2;L_\kappa)
 &=
 \int_0^\infty\df t\,
 e^{-\kappa t e^{-\gamma_E}}
 I_{ij}^Y(\bar z,\mu^2;t)\,,
\nn\\
 L_\kappa&=\ln\frac{\kappa\mu^2}{Q^2}\,.
\label{eq:laplace_transform_kernel}
\end{align}
In Laplace space, convolutions in $t$ become ordinary products.
The $\mu$-independence of \eq{single_collinear_factorization}, together with the evolution of the hard function and the PDFs, implies
\begin{align}
 \frac{\df}{\df\ln\mu^2}
 \widetilde I_{ij}^Y(\bar z,\mu^2;L_\kappa)
 &=
 -\sum_k
 \widetilde I_{ik}^Y(\bar z,\mu^2;L_\kappa)
 \otimes_z P_{kj}\bigl(z,\as(\mu^2)\bigr)
 \nn\\
 &\quad+
 \left[
 \Gamma_{\rm cusp}^{r_i}(\as)L_\kappa
 +\frac12\gamma_B^{r_i}(\as)
 \right]
 \widetilde I_{ij}^Y(\bar z,\mu^2;L_\kappa)\,.
\label{eq:kernel_RGE}
\end{align}
Our splitting-function convention is
\begin{align}
 \frac{\df}{\df\ln\mu^2}f_i(x,\mu^2)
 =\sum_jP_{ij}\bigl(x,\as(\mu^2)\bigr)\otimes_x f_j(x,\mu^2)\,.
\label{eq:DGLAP}
\end{align}
It follows from \eqs{beam_matching}{kernel_RGE} that the full rapidity beam
function obeys the multiplicative evolution equation
\begin{align}
 \frac{\df}{\df\ln\mu^2}
 \ln\widetilde B_i^Y(\bar\xi_2,\mu^2;L_\kappa)
 =
 \Gamma_{\rm cusp}^{r_i}(\as)L_\kappa
 +\frac12\gamma_B^{r_i}(\as)\,.
\label{eq:beam_RGE}
\end{align}
In particular, the anomalous dimension is independent of $\bar\xi_2$, as is
generally the case for leading-power beam functions~\cite{Stewart:2009yx}.

We expand the anomalous dimensions and splitting functions according to
\begin{align}
 \Gamma_{\rm cusp}^{r_i}(\as)
 &=\sum_{n=0}^\infty
 \left(\frac{\as}{4\pi}\right)^{n+1}\Gamma_n^{r_i}\,,
\nn\\
 \gamma_B^{r_i}(\as)
 &=\sum_{n=0}^\infty
 \left(\frac{\as}{4\pi}\right)^{n+1}\gamma_{B,n}^{r_i}\,,
\nn\\
 P_{ij}(z,\as)
 &=\sum_{n=0}^\infty
 \left(\frac{\as}{4\pi}\right)^{n+1}P_{ij}^{(n)}(z)\,.
\label{eq:anomalous_dimension_expansions}
\end{align}
The three-loop calculation requires $\Gamma_n^{r_i}$,
$\gamma_{B,n}^{r_i}$, and $P_{ij}^{(n)}$ for $n\le2$.
The cusp anomalous dimension and three-loop splitting functions are known from
\refscite{Korchemsky:1987wg,Moch:2004pa,Vogt:2004mw}.
The noncusp beam anomalous dimensions in the conventions used here are
collected in \refcite{Billis:2019vxg}.

\subsection{Structure of the rapidity beam-function counterterm}
\label{app:counter_term}

The hard counterterm $Z_H^{r_i}$ appearing in \eq{kernel_renormalization}  obeys the renormalization group equation 
\begin{equation}
\label{eq:ZH_anomalous_dimension}
\frac{\df\ln Z_H^{r_i}}{\df\ln\mu^2} = \Gamma^{r_i}_{\rm cusp}(\as)\,L_Q
 + \frac{1}{2}\gamma_H^{r_i}(\as)\,,
\end{equation}
where $\gamma_H^{r_i}$ is expanded as in \eq{anomalous_dimension_expansions}, with
$\gamma^{q}_{H,0}=-12C_F$ and $\gamma^{g}_{H,0}=-4\beta_0$.
With $\beta(\alpha,\eps)$ as defined in \eq{d_dimensional_beta} below, the
$\MSbar$ solution of \eq{ZH_anomalous_dimension} reads
\begin{equation}
\label{eq:ZH_all_order}
\ln Z_H^{r_i} = \int_0^{\as(\mu^2)}\frac{\df\alpha}{\beta(\alpha,\eps)}
\bigg[-4\Gamma^{r_i}_{\rm cusp}(\alpha)\int_{\as(\mu^2)}^{\alpha}
\frac{\df\alpha'}{\beta(\alpha',\eps)}
+2\Gamma^{r_i}_{\rm cusp}(\alpha)\,L_Q+\gamma_H^{r_i}(\alpha)\bigg]\,.
\end{equation}
The threshold component of mass factorization is transformed to Laplace space as
\begin{equation}
\label{eq:Gammas_laplace}
\widetilde\Gamma^{r_i}_s(\kappa,\mu^2,\eps) = \int_0^\infty \df t\,
e^{-\kappa t e^{-\gamma_E}}\,\Gamma^{r_i}_s(1-t,\mu^2,\eps)\,,
\end{equation}
where distributions in $t$ are extended to $t\in[0,\infty)$ with their
regularization boundary at $t=1$, as for $\cL_m(\bar\xi_2)$ in \eq{measurement_distribution}.
Since $\Gamma^{r_i}_s$ does not depend on $Q^2$, $\widetilde\Gamma^{r_i}_s$ depends
on $\kappa$ only through $\ln\kappa$, which combines with the $L_Q$ dependence
of $Z_H^{r_i}$ into the dependence on $L_\kappa$. Finally, writing the endpoint
behaviour of the diagonal splitting functions as
\begin{equation}
\label{eq:P_endpoint}
P_{ii}(z,\as) = \Gamma^{r_i}_{\rm cusp}(\as)\Big[\frac{1}{1-z}\Big]_+
 + P_\delta^{r_i}(\as)\,\delta(1-z) + \cO\big((1-z)^0\big)\,,
\end{equation}
the $\mu$ independence of \eq{single_collinear_factorization} fixes the beam anomalous dimension in
terms of the hard anomalous dimension and the splitting functions,
\begin{equation}
\label{eq:gammaB_relation}
\gamma_B^{r_i}(\as) = -\gamma_H^{r_i}(\as) - 2P_\delta^{r_i}(\as)\,,
\end{equation}
which gives, e.g., $\gamma^q_{B,0}=6C_F$ and $\gamma^g_{B,0}=2\beta_0$, in
agreement with \refcite{Billis:2019vxg}.

It is useful to combine the hard counterterm and the threshold component of
mass factorization into a Laplace-space beam counterterm,
\begin{align}
 \widetilde Z_B^{r_i}(L_\kappa,\mu^2,\eps)
 =Z_H^{r_i,-1}(Q^2,\mu^2,\eps)\,
 \widetilde\Gamma_s^{r_i}(L_\kappa,\mu^2,\eps)\,.
\label{eq:ZB_definition}
\end{align}
The renormalization formula \eqref{eq:kernel_renormalization} then takes the
compact form
\begin{align}
 \widetilde I_{ij}^Y(\bar z,\mu^2;L_\kappa)
 =
 \sum_k\widetilde Z_B^{r_i}(L_\kappa,\mu^2,\eps)
 \left[{\bf Z}_{\as}\,
 \widetilde I_{ik}^{Y,\mathrm{bare}}(\bar z,\eps;L_\kappa)\right]
 \otimes_z\Gamma_{kj}(z,\mu^2,\eps)\,.
\label{eq:kernel_renormalization_laplace}
\end{align}

The pole structure of $\widetilde Z_B^{r_i}$ follows directly from
\eqref{eq:beam_RGE}.
Our sign convention is fixed by
\begin{align}
 \frac{\df}{\df\ln\mu^2}\ln\widetilde Z_B^{r_i}
 =\Gamma_{\rm cusp}^{r_i}(\as)L_\kappa
  +\frac12\gamma_B^{r_i}(\as)\,.
\label{eq:ZB_anomalous_dimension}
\end{align}
Writing
\begin{align}
 \beta(\as,\eps)
 \equiv\frac{\df\as}{\df\ln\mu}
 =-2\eps\as+\beta(\as)\,,
\qquad
 \beta(\as)=-2\as\sum_{n=0}^\infty\left(\frac{\as}{4\pi}\right)^{n+1}\beta_n\,,
\label{eq:d_dimensional_beta}
\end{align}
with
\begin{align}
 \beta_0=\frac{11}{3}C_A-\frac43T_Fn_f\,,
 \qquad
 \beta_1=\frac{34}{3}C_A^2-\frac{20}{3}C_AT_Fn_f-4C_FT_Fn_f\,,
\label{eq:beta_coefficients}
\end{align}
where $C_A=n_c$, $C_F=(n_c^2-1)/(2n_c)$, $T_F=1/2$, $n_c$ is the number of colors, and $n_f$ is the number of light quark flavors; see e.g.\ \refscite{Tarasov:1980au,Larin:1993tp}.
One obtains
\begin{align}
 \ln\widetilde Z_B^{r_i}
 =\int_0^{\as(\mu^2)}
 \frac{\df\alpha}{\beta(\alpha,\eps)}
 \left[
 4\Gamma_{\rm cusp}^{r_i}(\alpha)
 \int_{\as(\mu^2)}^\alpha
 \frac{\df\alpha'}{\beta(\alpha',\eps)}
 +2\Gamma_{\rm cusp}^{r_i}(\alpha)L_\kappa
 +\gamma_B^{r_i}(\alpha)
 \right].
\label{eq:ZB_all_order}
\end{align}
The required anomalous-dimension coefficients are collected in \refcite{Billis:2019vxg}.
This counterterm predicts all poles of the bare kernels that are not removed by coupling renormalization and the non-endpoint part of mass factorization, and therefore provides a stringent check on the bare calculation.

\addcontentsline{toc}{section}{References}
\bibliographystyle{jhep}
\bibliography{refs_main}

\end{document}